# Cyclic Maxwell Demon in granular gas using

## 2 kinds of spheres with different masses

**P. Evesque**
Lab MSSMat,  UMR 8579 CNRS, Ecole Centrale Paris
**92295 CHATENAY-MALABRY, France,** e-mail: pierre.evesque@ecp.fr

**Abstract:**

*The problem of Maxwell's demon in granular gas is revisited in the case of a mixture of two particle species. The phase space is found to be 2d. Existence of cyclic orbits, with periodic segregation, is demonstrated by investigating the case of 2 kinds of particles with identical parameters but different masses. At large excitation equi-partition shall be obtained, but convergence towards the steady state is found in spiral. The spiral convergence is imposed due to the rule of kinetic-energy transfer between the two species. It results that the most probable scenario is that the steady state breaks into cyclic orbit at lower amplitude of vibration below a bifurcation threshold. The nature of  the bifurcation is not known; it can be critical, subcritical, hypercritical or can exhibit a tri-critical point as varying the control parameters. No conclusion is obtained at very low vibration amplitude: it is guessed two scenarii under further cooling which generates Maxwell's demon and segregation..*

**Pacs # : 5.40 ; 45.70 ; 62.20 ; 83.70.Fn**

Let us consider the following problem which is connected to the Maxwell's demon effect in granular gas [1-4]: be two adjacent vertical halve boxes (labelled 1 and 2 hereafter) connected by a horizontal slit at a given height h from the bottom; this slit has a given width w. The two boxes are assumed to be identical and shaken vertically; if the two boxes contain a single kind of particles, one gets the so-called experiment on "Maxwell's demon in granular gas".

Take now the possibility of having two species of grains, say X, Y, with different dissipation or size and shake the whole set-up vertically at frequency $f_r=\omega/(2\pi)$ and amplitude A. Is this problem of the same nature as the classic Maxwell's demon in granular gas? The response seems to be "yes" from the literature [5]. But is it true? How does it combine with a segregation effect for instance?

So, one should answer "no", just for safety, simply because the dimension of the phase space is increased due to the number of free unknowns. These unknowns are the numbers $N_{x1}$, $N_{x2}$, $N_{y1}$, $N_{y2}$ of particles in each box, and their time evolution, so that behaviour complexity is increased, and this should be observed in some range of parameters. But can one observe a chaotic behaviour? The aim of this paper is to try to characterise the complexity of the system.

In order to simplify the argumentation, the paper considers first the case when (X,Y) particles are spherical and identical, *i.e.* same radius $r_x=r_y$, same restitution coefficient, same solid friction, except that their masses $M_x$ & $M_y$ are assumed to be different; say $M_x>M_y$ . It considers also only the case when boxes are identical and contain a small number of layers and the slit width w is small, so that the operating conditions at work for a dilute granular gas are achieved, for which the flows $J_{1\to 2}$ and $J_{2\to 1}$ control the physics of particle migration. In fact, the paper will prove that one shall observe in some circumstances that the (X,Y) particles separate partly and that some periodic oscillation of population occurs; this means in particular that the content of each box shall not be steady but varies periodically with time.

Also, the paper demonstrates that the phase space dimension of the system is 2 when the experiment operates at constant total number of each kind of particles, as far as the vibration excitation can be considered as continuous. So the attractors cannot be chaotic, but they can be limit-cycles (*i.e.* indicating periodic evolution) or points (*i.e.* this last case indicates a steady state regime with or without segregation).

*Governing equations:* Be $x_1$ (or $y_1$) the number of grains of type X (or Y) at time t in box 1 and $x_2$ (or $y_2$) the number of grains of type X (or Y) in box 2. Be $f(x_1,y_1)$ the flux of beads X from one box (says 1) and $g(x_1,y_1)$ the flux of beads Y from the same box. Conversely, the flux from the other box will be given by the same functions f and g, *i.e.* $f(x_2,y_2)$ & $g(x_2,y_2)$ respectively, because of the symmetry of the system (the two containers are identical) and because of the assumed hypotheses, *i.e.* the granular medium is a low density gas and the slit is small. Be 2m and 2n the total number of particles X and Y respectively.

As the system is closed one gets:

$$x_1 + x_2 = 2m \tag{1}$$

$$y_1 + y_2 = 2n \tag{2}$$

$$dx_1/dt = -dx_2/dt = -f(x_1,y_1) + f(x_2,y_2) \tag{3}$$

$$dy_1/dt = -dy_2/dt = -g(x_1,y_1) + g(x_2,y_2) \tag{4}$$

One can change of variables and use the set u, v defined as $x_1=x_o+u$, $x_2=x_o-u$, $y_1=y_o+u$, $y_2=y_o-u$. Eqs. (1,2) are automatically satisfied, if $x_o=m$, $y_o=n$; Eqs. (3,4) become:

$$du/dt = f(m-u, n-v) - f(m+u, n+v) \tag{5.a}$$

$$dv/dt = g(m-u, n-v) - g(m+u, n+v) \tag{5.b}$$

Eqs. (5) is a set of two coupled linear differential equations with two unknowns (u,v). They are controlled by a set of control parameters among which are at least the amplitude A and frequency $f_r$ of vibration the total numbers 2n and 2m of X, Y particles, the height h of the slit, the shape of the box motion (sinus, triangle,…)…. In the theory of dynamical system [6], it would be said that the dimension of the phase space is 2. So this set cannot lead to chaotic behaviour; nevertheless it may generate cyclic attractors. The question is: when is it possible to get cyclic behaviour?

**Note on Chaos and strange attractor generation:**
If one assumes that chaos is found experimentally nevertheless, this would indicate that some of the above hypotheses are wrong. This point is discussed now:

One notices first that a modification of the flow functions is not able to make the problem chaotic because it will not change the number of equations and the number of unknowns: indeed, Eqs. (1) and (2) remain still valid and the phase space dimension remain equal to 2 if the modification wears only on the flow rules used in Eq. (5): Writing $f_1(x_1,y_1|x_2,y_2)$, $f_2(x_2,y_2|x_1,y_1)$, $g_1(x_1,y_1|x_2,y_2)$, $g_2(x_2,y_2|x_1,y_1)$ instead of $f_1(x_1,y_1)$, $g_1(x_1,y_1)$, $f_2(x_1,y_1)$, $g_2(x_1,y_1)$ does not change the number of unknowns.

On the other hand, if one observes chaos in such a closed experiment, it would mean that the phase space is equal to 3 at least [6]. In turn, it means that time becomes an important parameter, so that the excitation by vibration can not be considered as steady any more, but shall depend on time. This leads to flows depending on time. It may occur for instance when the time of flight of some particle at some time becomes small compared to the frequency of excitation.

**Steady excitation:**
We turn back now to the investigated case, for which excitation is considered as steady and flows depend on the content of the box from where the balls are leaving only. A question arises: Can one expect that a steady solution exists in some range of working parameters. This steady solution means a point attractor, *i.e.* $(u_1,v_1)$.

*First result:* According to Eq. (5), u=v=0 is always a steady solution since it leads to du/dt=dv/dt=0. However it remains to study the stability of this solution. In fact, this solution corresponds to equi-repartition. So, *a priori*, this solution shall be found at large excitation and small number of balls. This is at least what is observed in the case of "true Maxwell's demon effect in granular gas" [1-4].

Then we shall start from this solution to investigate further, and find more complex situations. So, we start from (u=0,v=0) and study the evolution of this steady solution when changing the control parameters, and we look for solutions in a region where u and v remain small compared to m and n. Also we consider cases when functions f and g are smooth enough so that first derivative of f and g will be sufficient to take account of the dynamics. Of course, in the case when a bifurcation occurs, this approximation (of smallness of u and v) is not achieved anymore when the undergone bifurcation is of sub-critical nature, or when the system is driven far away from the bifurcation threshold. This will not be studied because it requires knowing the general trend of the functions f & g everywhere in the (u,v) plane, which is not known at the moment. However, in most cases of Maxwell's demon experiment in granular gas (with a single kind of particles) a critical bifurcation is observed. So, we will guess that same situation may occur here, for which (u,v) remains small near the threshold. Nevertheless, we will look for situations more complex here, because the attractor which we try to detect may be periodic. In other words, we want to know if it is possible to see the content of the 2 boxes evolving periodically, with some balls filling/emptying the boxes alternately and periodically.

So, the first step in analysing this problem is to perform a first-order expansion of Eq. (5) or a "linearization". One has to introduce two derivatives for each flows f and g. We use the following notations $f'=f_x =\partial f(x,y)/\partial x$ ($g'= g_x=\partial g/\partial x$) which corresponds to X grains, and $f''=f_y =\partial f/\partial y$ (resp. $g''=g_y=\partial g/\partial y$) which corresponds to Y grains. In other words, this corresponds to: $f'=f_x(m, n)$; $f''=f_y(m, n)$; $g'=g_x(m, n)$; $g''=g_x(m, n)$.
So, $f(m-u,n-v)= f(m,n) -uf' -vf''$, and so on.
Eq. (5) writes:

$$du/dt = -2uf' - 2vf'' \tag{6.a}$$

$$dv/dt = -2ug' - 2vg'' \tag{6.b}$$

So Eqs. (6) govern the evolution of populations at first order expansion:

For the steady state $u=0 = v=0$ be a stable steady state, one needs that the system of equations gets two eigen values with negative real part each. This ensures the convergence of the dynamics of any perturbation to $u=v=0$. The eigen values are solution of the characteristic equation :

$$(\lambda+2f')(\lambda+2g'') - 4f''g' = 0 = \lambda^2 + 2\lambda(f'+g'') + 4[f'g'' - f''g']$$

So defining

$$\Delta=(f'+g'')^2 + 4f''g' - 4f'_xg''=(f' -g'')^2 + 4f''g' \tag{7}$$

One gets the two eigen values $\lambda_\pm$:

$$\lambda_\pm = -[f'+g'' \pm (\Delta)^{1/2}] \tag{8}$$

We are concerned in determining whether the experiment can produce cyclic behaviour or not. Indeed getting oscillation needs complex eigen values $\lambda_\pm$, which requires in turn that $\Delta$ is negative, since $f'$ and g" are real. Assuming then a negative $\Delta$, the system, when perturbed, will converge in spiral towards $u=v=0$ if $f'+g'' >0$, while it will diverge from it when $f'+g'' <0$. So the conditions of a Hopf bifurcation [6] occurs when $(f'+g'')$ changes of sign, while $\Delta$ is negative typically. We recall in appendix the generic equation of the Hopf bifurcation and its subcritical/critical nature as a function of its parameters. We develop there also Eq. (5) at higher order to write the parallel with Hopf bifurcation in more details. However, it is difficult to determine all these terms theoretically so that discussion about the true critical/subcritical nature of the Hopf bifurcation cannot be done further. So we discuss further the physics of the phenomena with Eqs. (6-8) in the next paragraph, and we focus on $\Delta$.

**Parameters of the dynamics**

*1-    large excitation range*

    *or the limit of non interacting particles:*

In the limit of small m and n, particles do not interact. This shall ensure *a priori* equi-repartition, since flows from the slit shall increase linearly with the ball number in the box in this case. This reads f(x)=kx and g(y)=ky. This ensures the stability of the solution u=v=0 in the case of two coupled boxes, since one gets f' =g">0, f"=g'=0. But we are looking now to this point in more detail.

*Recall:* if we consider the result from Maxwell's demon [1-4] with a single kind of ball, equi-repartition occurs when there is little number of particles in each box and/or when excitation is large enough and that ball number N is small enough. In this case, it is known now [2-3] that the flow J at a given set $(A,f_r)$ of vibrational parameter is found to increase linearly with the grain number N; then it saturates, reaches a maximum at $N_m$ and decreases with increasing N further. This is exemplified in Fig. 1.

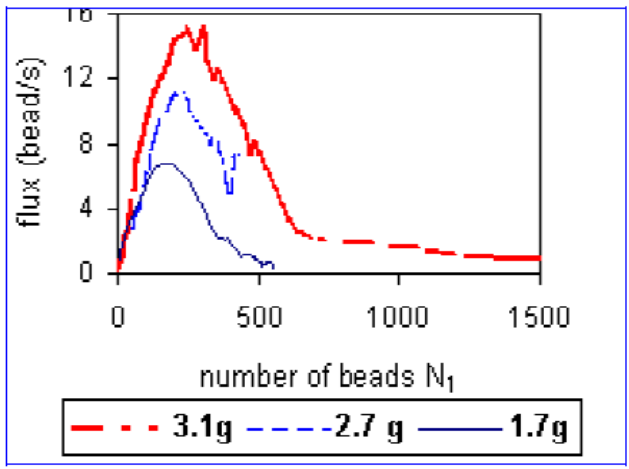

*Figure 1:  The flow of beads from a box with a slit, as a function on the number* N *of grains in the box in the case of a single kind of grains, at different excitation acceleration (from [2]).*

Still in the case of a single kind of balls, equi-repartition is ensured till the maximum [2]. In the first linear regime the flow is proportional to N, so everything occurs as if the dynamics of each ball was independent from the others [7].

We consider now the case of a mixture of 2 kinds of balls; we first note that the dynamics of N identical balls with same mass is independent of the particle mass [8] (in term of speed and not in term of momenta or energy) . This is due to the equality between inertial mass and weighting mass. So, if we consider now two sets (X,Y) of balls which differ only from their mass, *i.e.*  $M_X \neq M_Y$, $M_X > M_Y$, these two sets shall exhibit similar dynamics. So both sets shall exhibit similar ball speed distribution. So, in this limit regime of small ball numbers, one would expect that

$\quad$ f"=g'=0, $\hfill$ (9.a)

$\quad$ f'=g"=$k_o$>0 , $\hfill$ (9.b)

Eq. (9.a) indicates that each particle do not interact with the others and Eq. (9.b) that the flow is proportional to the number of particles in the box; so the number of particles X flowing from the box is just proportional to x, and independent of y, and conversely for Y. This is just what Eq. (9) tells.

Eq. (6) is simple to solve in this case upon these values for f', f", g', g" . One finds the steady solution:

$$u=v=0 \qquad (7)$$

and this solution is found to be stable. So, this analysis forecasts an equi-repartition as we were expecting.

## *2- Intermediate regime:  Intermediate excitation range*
### *Or intermediate range of ball numbers  2m and 2n:*

When m and n are increased, interactions between balls appear. This generates dissipation. The first effect of this dissipation is to reduce the values of the derivative (as in Fig. 1):

$$f' >0, \text{ but } f'<k_o , \text{ and } f' \text{ decreases when m increases} \qquad (8.a)$$

$$g" >0, \text{ but } g"<k_o , \text{ and } g" \text{ decreases when m increases} \qquad (8.b)$$

We know also from Maxwell's demon experiment that further increase of the numbers n and m may generate negative f' and g" [2] , see Fig.1.

On the other hand, different trends happen for f" and g'. Indeed, both kinds of beads do not wear the same kinetic energy when they have the same speed, because they have different masses. So following an argument developed in [8], their collisions allow some transfer which tends to equilibrate their momenta and their kinetic energy, instead of imposing similar speed distribution. Hence what occurs in a XY collision in average is that the ball which has the less mass, *i.e.* Y, will acquire some larger kinetic energy and the one which is the heavier, *i.e.* X, will loose some energy to the benefit of the other one. Hence, one expects :

$$f" <0, \text{ and } f" \text{ decreases from 0 when m or n increases} \qquad (8.c)$$

$$g' >0 , \text{ and } g' \text{ increases from 0 when n or m increases} \qquad (8.d)$$

Also, g' (and f") start increasing (decreasing) from 0 when increasing m and n from 0.

Combining these trends with the population dynamics (Eqs. 6), leads to a negative $\Delta=-\Omega^2$, hence to complex eigen values: indeed, $\Delta=(f'+g")^2 + 4f"g' - 4f'g"=(f'-g")^2 + 4f"g'$ from Eq. (7); so assuming f'=g", one gets that $\Delta<0$. In other words, convergence to equi-distribution shall occur in spiral as soon as the flow starts behaving not linearly with the ball number.

So starting with small m and n numbers of particles, (and high excitation) f'=f'$_o$ ≈ g'=g'$_o$ and f''=f''$_o$≈ 0, g' ≈ 0 ; in this case the state with equi-partition is stable since the real parts of its eigen values are approximately –f'$_o$ and –g''$_o$, so that they are negative. In fact, f'$_o$ and –g''$_o$, are expected to depend on the excitation speed aω.

Increasing the number of balls f' and g'' start deceasing, but remains equal about (f' = g''); however f'' becomes negative and g' positive so that Δ becomes negative, implying complex eigen values λ$_±$= –(f' +g'')/2±iΩ) = α$_±$ ±iΩ with Ω²= –Δ. So convergence towards u=v=0 is in spiral.

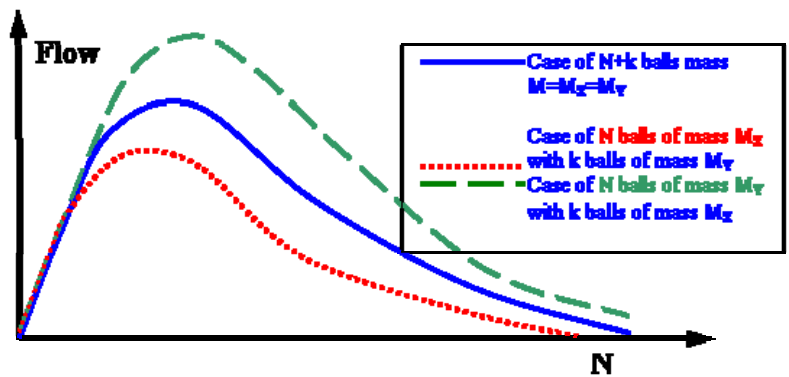

***Figure 2 :*** *Typical flow curves obtained in a box containing mixtures of heavy (X) and light (Y) particles. The number N of X (resp. Y) particle is varied at fixed Y (resp. X) particles. The flow red (resp. green) curves correspond to the flow from the N particles X (resp. Y). The blue curve correspond to the ratio N/(N+k) of the flow obtained with a box containing N+k identical particles (either X or Y).*

## *3- General trend of flow rules:*

Let us now consider the case of a box containing the two species X,Y, one with the amount N and the other with a fix amount k. From what has been told previously, it is obvious that the flow curve of the lighter shall be improved by the presence of the heavier, while the flow of the heavier shall decrease. It results from this that we expect the typical flow curves as those reported in Fig. 2.

In this Figure, the blue (continuous) curve represents the flow J$_{N,k}$ from a subset of N particles in a box containing N+k identical particles (either the lighter or the heavier); it is given by J$_{N,k}$ =J(N+k) N/(N+k) , where J(N+k) is the flow from a box containing N+k identical particles. The green (dash) curve is the one for N light particles in presence of k heavier particles, while the red (dots) curves correspond to the flow of N heavy particles with k lighter. This Figure was built according to the preceding rules: Mixing N light particles with k heavier shall improve the mean speed of the heavier and decrease the mean speed of the lighter. From this also, one shall conclude that the distance from the blue-red (*resp.* blue-green) curves shall depend on k : The larger the k the larger the difference (still in the limit of small k). Also, one shall conclude from this figure that the positions N$_{max,X}$ & N$_{max,Y}$ of the curve maximum (in N) depend on the ball mass M$_X$ (or M$_Y$), on their difference (M$_X$ –M$_Y$) and on k. At last, f' and g'' correspond to the derivative of the red and green curves with respect to x and y respectively; similarly f" and g' can be computed from the change of these curves as a function of k.

According to these statements, the Figure indicates that f' can be negative before g". Estimates of f" and g' are obtained under linear approximation from Figure 2 by subtracting the values of blue-red & green-blue flows at a given N, followed by division by k. From this, one expects f" g' <0. We first look at the instability of the steady state (u=v=0), then look at larger distance from this bifurcation. This allows sketching the behaviour evolution as a function of the excitation parameter aω.

## *4- Sketch of evolution of behaviour:*

### *A- Small-Amplitude Oscillating regime and beyond: Hopf bifurcation:*

Dynamics of population is controlled by Eq. (6) and the solution of eigen values and eigen states. When excitation is large, the steady state is (u=v=0) and convergence is exponential with a constant time $1/(f'+g")$. A convergence in spiral towards this point shall be observed at an oscillation rate $\Omega_o$, such as $\Omega_o^2=-\Delta$, which starts with a zero frequency, since f"g'=0 when v=aω is large, then increases with the increase of -f"g'. But above some threshold $V_o=a_o\omega_o$, which corresponds to f'+g"=0, the steady state (u=v=0) is no more stable and a bifurcation occurs. This forces the system to oscillate with finite amplitude ε at the frequency $\Omega^2=-\Delta$ given by Eq. (7).

A question arises: what is the variation of the amplitude of oscillation with the distance to the bifurcation: in classic Hopf bifurcation symmetry considerations impose that the second order term in the perturbation is 0. Here too, as demonstrated by Eq. (A5) of the appendix, in which the coefficient of the second order term is 0 due to the symmetry of the flows. Since some of the coefficients of third order term are likely not 0, then assuming it to be positive, hence one expects an amplitude ε of oscillation which scales as $\varepsilon \propto (V_o-V)^{1/2}$. So:

$$\varepsilon \propto (V_o-V)^{1/2} = (a_o\omega_o-a\omega)^{1/2} \qquad (9.a)$$

$$\Omega = (-\Delta)^{1/2} \qquad (9.b)$$

So this bifurcation leads to generate a periodic oscillation of population, which starts above the threshold $a_o\omega_o$ with a finite frequency $\Omega_o$ ($\Omega_o \neq 0$) and which varies slowly; meanwhile the amplitude ε of population oscillation starts at 0 at $a_o\omega_o$. and increases as $\varepsilon \propto (a_o\omega_o - a\omega)^{1/2}$, when decreasing the parameter of vibration.

• *Note:* nothing imposes the third order terms to be positive, or the fifth order terms… So the bifurcation can also be of the sub-critical type. Also these terms can depend on the parameters a and ω independently, on other parameter 2m,…, so that a tri-critical point can appear as in [4bis], or one can meet a hyper-critical bifurcation case [4].

### B- *large amplitude oscillation and Beyond oscillations:*

Far after this bifurcation and continuing decreasing the vibration energy, the discriminant Δ evolves. As the functions f and g are quite non linear, Δ may become positive which will lead to a stop of the oscillation of population. However, if this happens, the solution ($x_1 = x_2 = m$, $y_1 = y_2 = n$) will be likely unstable because of the sum f'+g'' which will be negative, so that one expects in turn a bifurcation toward a Maxwell's Demon steady state (MD) at smaller excitation. This MD state can be segregated, *i.e.* the composition of the two boxes may be different and their filling different too.

However the existence of a stable segregated steady state is not linked to the fact Δ can become positive; it results from the negativity of f' & g'' at x>m & y>n respectively for small enough excitation, so that cyclic behaviour cannot be stabilised perhaps for a while by the coupling f'' and g'.

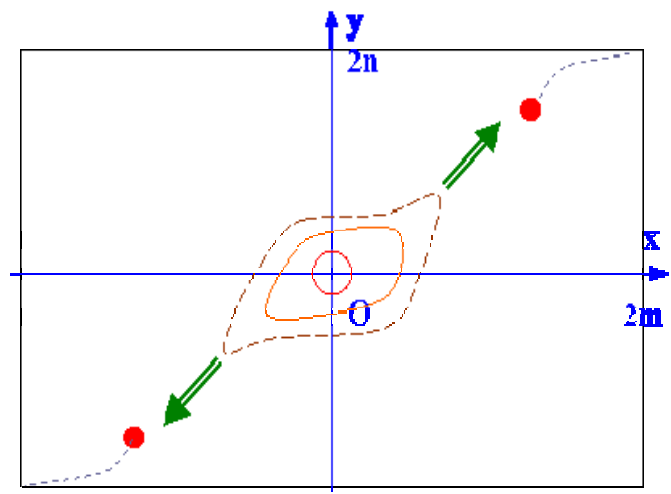

*Figure 3: Possible scenario of evolution: here are represented different possible trajectories of the system in the (x,y) plane at different amplitudes of vibration. At large excitation, the attractor is a point: the point O ($x_1=x_2 =m$, $y_1=y_2 =n$). Below some threshold, a Hopf bifurcation occurs, and the attractor is a closed loop indicating a cyclic (red circle) whose size expands and shape deforms when decreasing the amplitude of vibration (orange). At some other threshold (dashed brown), this trajectory meets a saddle point and the system escapes and falls in one of the two states of Maxwell's demon (red points) (MD points). The MD points evolve towards the bottom-left and top-right corners when lowering excitation*
***Question:*** *is this script of evolution reversible starting from the Maxwell demon state, when increasing the Amplitude.*
*The two red points correspond to Maxwell's demon attractors (MD); their trajectory is indicated in dashed grey when decreasing the amplitude of vibration.*

Anyhow, the steady state which appears at smaller vibration shall exhibit large difference of populations so that the analysis in the basis of equi-distribution is not correct. The best way to study this new trend is to reconstruct the different possible attractors at different stages of excitation using the phase space of the system. This is done in Fig. 3. Indeed the variations of functions f and g are expected to be so large in this phase space so that a dynamics study using classic expansion method in the vicinity of ($x_1 = x_2 = m$, $y_1 = y_2 = n$) is meaning less anymore. A possible expansion in the vicinity of a MD steady state is still possible but difficult because no one knows the values of $f'(x_1,y_1)$, $f'(x_2=m-x_1,y_2=n-y_1)$, $f''(x_1,y_1)$, $f''(x_2,y_2)$, $g'(x_1,y_1)$, $g'(x_2,y_2)$, $g''(x_1,y_1)$, $g''(x_2,y_2)$, at this working point and their possible evolution when increasing the excitation parameter $a\omega$.

So, when the population of the system stops evolving periodically, it freezes in some configuration which exhibits a non equi-partition in both compartments. This means $x_1 \neq x_2$ and $y_1 \neq y_2$. However, due to symmetry considerations it is obvious that the symmetric solution which interchanges box indices 1 and 2 is also a steady solution. We have represented these two solutions by the two red points in Fig.3 in the top right and bottom left quarters. One expects that these attractors evolve towards the left bottom corner and top right corner when excitation decreases, since these locations correspond to $(x_1=y_1=0)$, or $(x_2=y_2=0)$. We need a scenario to understand what is occurring before this stage.

We have also represented the cycles in Fig. 3, obtained after the Hopf bifurcation at $V_o=a_o\omega_o$; their trajectories shall turn around the point O of coordinates $(x_1=x_2=m, y_1=y_2=n)$, since O corresponds to equi-repartition and since oscillations are generated from a bifurcation from this attractor point. Further decrease of the excitation increases the size of the trajectory and deforms it. In the vicinity of the Hopf bifurcation (at point O), the x direction is unstable and the y is stable (see Fig. 3), and (x,y) coupling makes the system rotating. Decreasing the excitation sufficiently makes also the y direction unstable because it translates the maxima of Fig. 3 towards left. So at sufficiently slow excitation one expects that both Maxwell's demon effect on x and y species win and that trajectory ends at the red attractor points of Fig. 3.

To do so there may be two possible explanations: in the first one the trajectory may pass near a saddle point whose altitude decreases when decreasing the excitation; so the system may go through the neck towards the MD attractor at some low excitation threshold $V_1=A_1\omega_1 << V_o=A_o\omega_o$. Such a neck, if it exists, is probably located in between O and the MD attractor. The second explanation considers that the cyclic trajectory has to pass an edge oriented approximately perpendicular to the first bisector which is merely the line joining the 2 MD attractors and passes through point O. The altitude of the edge increases when lowering the excitation, so that the system cannot pass over the edge at some excitation level $V_1=A_1\omega_1$; and the symmetry breaks. The careful study of the trajectory in this phase space may allow determining which process is involved. The pulsation $\Omega$ of oscillation is also expected to decrease toward 0 when decreasing $A\omega$ towards $V_1$, since a point with very low speed shall be reached when approaching the neck or at the edge.

### *C- Are MD attractors steady states or oscillating states?*

Let us now consider the "Maxwell's demon" solutions and study their stability, their possibility of oscillations…. So we consider a possible MD steady state $(x_{1o}, x_{2o}, y_{1o}, y_{2o})$, such as $2m= x_{1o}+ x_{2o}$, $2n= y_{1o}+ y_{2o}$, with $x_{1o}< x_{2o}$ & $y_{1o}< y_{2o}$, and a state $(x_1, x_2, y_1, y_2)$ which is slightly apart from this position, say $(x_1, x_2, y_1, y_2)= (x_{1o}+u, x_{2o}-u, y_{1o}+v, y_{2o}-v)$. The equation of evolution (Eqs. (3,4)) writes:

$$du/dt = -f(x_{1o}+u, y_{1o}+v) + f(x_{2o}-u, y_{2o}-v) \tag{10.a}$$

$$dv/dt = -g(x_{1o}+u, y_{1o}+v) + g(x_{2o}-u, y_{2o}-v) \tag{10.b}$$

Also as $(x_{1o}, x_{2o}, y_{1o}, y_{2o})$ is a steady state, one gets:

$$0 = -f(x_{1o}, y_{1o}) + f(x_{2o}, y_{2o}) \tag{10.c}$$

$$0 = -g(x_{1o}, y_{1o}) + g(x_{2o}, y_{2o}) \tag{10.d}$$

Developing Eqs. (10.a-b) at first order in the vicinity of u=v=0, noting
$f'_1 = df/dx(x_{1o}, y_{1o})$, $f''_1 = df/dy(x_{1o}, y_{1o})$, $f'_2 = df/dx(x_{2o}, y_{2o})$, $f''_2 = df/dy(x_{2o}, y_{2o})$,
$g'_1 = dg/dx(x_{1o}, y_{1o})$, $g''_1 = dg/dy(x_{1o}, y_{1o})$, $g'_2 = dg/dx(x_{2o}, y_{2o})$, $g''_2 = dg/dy(x_{2o}, y_{2o})$
one gets:

$$du/dt = -u[f'_1 + f'_2] - v[f''_1 + f''_2] \tag{11.a}$$

$$dv/dt = -u[g'_1 + g'_2] - v[g''_1 + g''_2] \tag{11.b}$$

The stability of this solution is studied by determining the eigen values $\lambda_1 \pm$ of these equations, which are solutions of:

$$0 = [\lambda + f'_1 + f'_2][\lambda + g''_1 + g''_2] - [f''_1 + f''_2][g'_1 + g'_2]$$

or:

$$0 = \lambda^2 + \lambda[f'_1 + f'_2 + g''_1 + g''_2] + (f'_1 + f'_2)(g''_1 + g''_2) - (f''_1 + f''_2)(g'_1 + g'_2)$$

Finding $\lambda$ requires computing the discriminant $\Delta_1$:

$$\Delta_1 = [f'_1 + f'_2 - g''_1 - g''_2]^2 + 4(f''_1 + f''_2)(g'_1 + g'_2) \tag{12}$$

This leads to:

$$\lambda_{1\pm} = -\{f'_1 + f'_2 + g''_1 + g''_2 \pm [\Delta_1]^{1/2}\}/2 \tag{13}$$

One expects that the steady state $(x_1 < x_2, y_1 < y_2)$ corresponds to well developed MD, for which the population $x_1$ and $y_1$ are quite small, for which the flow **is linear with x and y**. This reads: $f'_1 = g''_1 = k_o > 0$ and $f''_1 = g'_1 = 0$ (see Eq. (9)). On the other hand, $f'_2 <$, $g''_2 < 0$. It results from this some difficulty to predict the sign of $\Delta_1$ and of $f'_1 + f'_2 + g''_1 + g''_2$. In particular it is known from experiment on Maxwell's demon with a single kind of grain that the sum $f'_1 + f'_2$ and $g''_1 + g''_2$ can change of sign when increasing the excitation parameter $A\omega$ at constant $A$ [4] while the product $f'_1 * f'_2$ and $g''_1 * g''_2$ remain negative. So it is difficult to argue further. Let us then state simply that the solutions defined by Eq. 13 are (i) stable and steady if the real part of $-\{f'_1 + f'_2 + g''_1 + g''_2 \pm [\Delta_1]^{1/2}\}/2$ is negative and (ii) are oscillating if $f'_1 + f'_2 + g''_1 + g''_2$ is negative and $\Delta_1$ be negative.

Let us now consider such a case of oscillation with MD for which $f'_1+f'_2+g''_1+g''_2$ and $\Delta_1$ are both negative and let us study if the system can stop oscillating when changing $A\omega$. In fact there are two ways:

(i) either the system undergoes a second Hopf bifurcation for which $\{f'_1+f'_2+g''_1+g''_2\}/2$ passes from negative to positive at some stage; in this case the amplitude of oscillation shall tends progressively to 0 and the oscillation pulsation shall remain merely constant $\Omega_1 = [-\Delta_1]^{1/2}$.

(ii) or the discriminent $\Delta_1$ passes from negative to positive at some stage. In this case, oscillation stops when pulsation goes to 0.

However an other case can merge if some of the 2 functions $f(x,y)$ or $g(x,y)$ can become 0 above some finite grain number (X or Y respectively) below some $A\omega$. This condition may become valid only during some part of the cycle. If so, one of the box empties completely of X or Y (depending on which function f or g gets 0), and the problem loose one degree of freedom. In such a case, one expects then to observe classic Maxwell demon for the second compound. The freezing and emptying of the heaviest particles seems to be the more likely owing to the rules for kinetic energy transfer between the balls. It means in this case that the system travels along the cyclic trajectory at a not constant speed, so that it can occur in some part of the cycle that speed goes very slow ($\rightarrow 0$). This will be this slow part of the trajectory which will control mainly the period of oscillation just above the quenching.

## 5- Generalisation to other systems and Conclusion:

We have considered mixtures of two sets of balls with same mechanical characteristics except from their mass. In fact as stated at the beginning of the paper the problem remain constraint mainly by its maximum complexity, which is determined by the dimension of its phase space. This one remains 2-dimensional as far as the excitation can be considered as continuous. Hence the complexity of the behaviour of the system cannot be larger than periodic orbit (no chaotic behaviour).

Also the transfer of kinetic energy from one kind of ball to the other one is mainly imposed by the difference of mass between the two species; this determines the sign of the coupling f"g' <0 in Eq. (6), which is negative; this leads in turn to a negative discriminant and to complex eigen values. Hence this forces oscillation and the generation of periodic obit. Indeed such oscillations have been found experimentally [9].

In turn, the coupling mechanism which allows oscillation shall apply also on other sets of 2 kinds of balls, as far as their masses are different. So, one shall expect that the trends described here are quite general and can be applied to most other cases. It means that one may expect that large amplitude vibration shall lead to steady equi-repartition; then the decrease of excitation shall generate a Hopf bifurcation with an oscillation of populations. This one stops in turn at smaller amplitude and the system undergoes a bifurcation towards a Maxwell's demon state (MD) with segregation. Different scenarii have been proposed for reaching this last stage.

Also possible evolution toward a cyclic behaviour with trajectory forming a loop around the MD state has been considered; since the transfer functions f and g are not known, this evolution is possible (in principle). In this case, further decrease of vibration amplitude will impose a second Hopf bifurcation to freeze the system into a steady state.

The nature of the Hopf bifurcation at high vibration amplitude has been rapidly discussed. Due to the uncertainty of the variations of the flow rules at large number of grains, one cannot predict exactly if the bifurcation will be critical or sub-critical; Considering the "true" Maxwell's demon case, at work with a single kind of balls, and its complexity, we may guess also such a complexity here, and assume (?) that the bifurcation can evolve and pass from critical to sub-critical via a tri-critical point.

It is possible that under some circumstances, oscillations and Hopf bifurcation do not occur so that the system undergoes directly some kind of steady "Maxwell's demon effect" as in classic granular gas, (but with segregation here). However it seems that it should be rare, since the coupling f'g"<0 seems to be logic as soon as particles have different masses: at similar speed the lighter particles have less energy and momenta than the heavier so that collisions between heavy-light particles will transfer kinetic energy from one species (the heavier ones) to the other/lighter ones.

At last, when dealing with particles of different sizes, it is likely important to think in terms of the number of layers the species covers on the bottom box at rest, instead of particle number, because this is what controls the typical mean free path and the collision losses.

As a matter of fact, studying the evolution of the content of two coupled boxes is a first step for understanding spatio-temporal complexity in a continuous space, since the system integrates both the coupling rules between adjacent locations and rules for time evolution of each species. So this study is a first step in understanding the space-time dependence of the patterning in segregation processes with two grain species. It is found that oscillation shall spontaneously occur with two boxes and two compounds only. So it may explain why segregation oscillation pattern can be seen in 3d segregation problem.

Furthermore, we can extrapolate this approach, to obtain some rule similar to the phase rule in thermodynamics that relates the number k of freedom degrees to the number n of compounds and the number $\varphi$ of different phases, which writes k=n+2-$\varphi$. Here, it becomes d=q(K-1) which relates the dimension of phase space d, *i.e.* the real number of unknown, to the number of grain species q and to the number of boxes K, since there are qK populations and a set of q coupled equations. So, as soon as d≥2 the system can oscillate under some condition, while it can exhibit chaos as soon as d≥3.

Applying this result to K boxes (K>2) and a single class of grains [8,9], this forecasts that 3 boxes may exhibit non steady state, but oscillating state while chaotic behaviour can be obtained when K>4 if transfer rules let do it. This does not seem to have been found nor envisaged [9-10] . In larger number of boxes one may expect generating system of wave propagations, solitons…. However, this requires specific transfer rules with large non linearity, which may be difficult to achieve with a single class of grains. Perhaps to include different slits at different height as in [9] may be sufficient? Anyway such behaviours shall likely be observed in the case of a larger number of boxes and compounds.

## Appendix: Hopf's Bifurcation:

The generic form of the dynamics of a system which undergoes a Hopf bifurcation writes:

$$dz/dt = (-\sigma - i\omega) z - z [\sum_k h_k |z|^{2k}] \quad \text{With } h_k = h'_k + ih''_k \quad \text{(A1)}$$

where $z = z_r + iz_i$ is a complex variable of two variables $z_r$ and $z_i$. We write $z = Z \exp(i\varphi)$, so that $Z = |z|$, we get

$$dZ/dt + iZ d\varphi/dt = (-\sigma - i\omega)Z - (h'_k + ih''_k)Z^{2k+1} \quad \text{(A2)}$$

so that

$$dZ/dt = (-\sigma - h'_k Z^{2k})Z \quad \text{(A3.a)}$$

$$Z \, d\varphi/dt = Z(-\omega - h''_k Z^{2k}) \quad \text{(A3.b)}$$

$z=0$ is a steady state. Since $\sigma$ and $\omega$ are real, $i\omega$ is imaginary and the eigen values $\lambda_\pm = -\sigma \pm i\omega$ in the vicinity of $z=0$ are complex. It results from this that convergence towards $z=0$ is ensured as long as $\sigma$ is positive, and that this convergence is in spiral. So, one can limit Eq. (A1) to its expansion at first order as long as $\sigma > 0$ because the system remains in the vicinity of $z=0$.

On the other hand, this solution $z=0$ is no more stable if $\sigma$ becomes negative. In this case, one shall take into account the higher order terms; be $2k+1$ the first one, for which we assume a positive real part $h'_k > 0$. Eq. (A3.a) has now the solutions when $\sigma$ is negative: the steady solution $Z=0$, which is now unstable when $\sigma$ is negative, and the solutions $Z = (-\sigma/h'_k)^{1/(2k)}$ when $\sigma < 0$ and $h'_k > 0$. Substituting this value in Eq. (A3.b), one obtains that the phase of z rotates linearly with time, since $d\varphi/dt = -\omega + \sigma/(h'_k h''_k)$, which indicates a periodic solution. In particular when $k=1$, one gets a classic supercritical bifurcation if $h'_1 > 0$ (and is stable); its amplitude grows rapidly with $\sigma$ in the vicinity of $\sigma = 0$; this generates important fluctuations. When $k > 1$, the sensitivity to $\sigma$ is even larger (since $Z = (-\sigma/h'_k)^{1/(2k)}$) so that fluctuations are more important even.

When the first $h'_k$ is negative, one has to expand further Eq. (A2) to include higher order terms; the bifurcation becomes sub-critical in this case.

The passage from this generic form to the case studied in the present article is straight forward: Eq. (A1) shall describes the dynamics of population which is given by Eq. .5. So, $z = u + iv$ and the pairs $(\sigma, \omega)$ & $(h', h'')$ are related to the expansion of functions f and g around m and n in Eq. (5a & b). The identification imposes at third order:

$$f(m-u, n-v) - f(m+u, n+v) = -\sigma u + \omega v - (u^2 + v^2)(h'u - h''v) \quad \text{(A4.a)}$$

$$g(m-u, n-v) - g(m+u, n+v) = -\sigma v - \omega u - (u^2 + v^2)(h'v + h''u) \quad \text{(A4.b)}$$

On the other hand, labelling $\partial f/\partial x = f_x$ …. and expanding f et g in the vicinity of m & n lead to :

$$f(m+u, n+v) = f + (uf_x + vf_y) + (u^2 f_{x^2}/2 + uv f_{uv} + v^2 f_{v^2}) + (u^3 f_{xxx} + 3u^2 v f_{x^2y} + 3uv^2 f_{xyy} + v^3 f_{yyy})/6 + (u^4 f_{xxxx} + 4u^3 v f_{xxxy} + 6u^2 v^2 f_{x^2y^2} + 4uv^3 f_{xyyy} + v^4 f_{yyyy})/24 + (u^5 f_{xxxxx} + 5u^4 v f_{xxxxy} + 15u^3 v^2 f_{xxxyy} + 15u^2 v^3 f_{xxyyy} + 5uv^4 f_{xyyyy} + v^5 f_{yyyyy})/120 + \ldots +$$

or

$$f(m-u,n-v) - f(m+u,n+v) = -2(uf_x + vf_y) - (u^3 f_{xxx} + 3u^2 v f_{x^2y} + 3uv^2 f_{xyy} + v^3 f_{yyy})/3 - (u^5 f_{xxxxx} + 5u^4 v f_{xxxxy} + 15u^3 v^2 f_{xxxyy} + 15u^2 v^3 f_{xxyyy} + 5uv^4 f_{xyyyy} + v^5 f_{yyyyy})/120 + \ldots$$

And similarly for g. So, this leads to the following dynamics equation:

$$du/dt = -2(uf_x + vf_y) - (u^3 f_{xxx} + 3u^2 v f_{x^2y} + 3uv^2 f_{xyy} + v^3 f_{yyy})/3 - (u^5 f_{xxxxx} + 5u^4 v f_{xxxxy} + 15u^3 v^2 f_{xxxyy} + 15u^2 v^3 f_{xxyyy} + 5uv^4 f_{xyyyy} + v^5 f_{yyyyy})/120 + \ldots + \quad \text{(A5.a)}$$

$$dv/dt = -2(ug_x+vg_y) - (u^3g_{xxx}+3u^2vg_{x^2y}+3uv^2g_{xyy}+v^3g_{yyy})/3 - (u^5g_{xxxxx}+5u^4vg_{xxxxy}+15u^3v^2g_{xxxyy}+15u^2v^3g_{xxyyy}+5uv^4g_{xyyyy}+v^5g_{yyyyy})/120 + \ldots + \quad (A5.b)$$

Identification imposes writing Eqs (A5) as the generic form (Eq. A4). Starting the identification with first order, one gets (see Eqs 7 and 8):

$$z=u+iv,\ \sigma = f'+g'' = f_x + g_y \text{ and } -\omega^2 = \Delta \qquad (A6)$$

Then limiting the analysis at third order, one gets:

$$du/dt = -2(uf_x+vf_y) - (u^3f_{xxx}+3u^2vf_{x^2y}+3uv^2f_{xyy}+v^3f_{yyy})/3 - \ldots + \qquad (A7.a)$$

$$dv/dt = -2(ug_x+vg_y) - (u^3g_{xxx}+3u^2vg_{x^2y}+3uv^2g_{xyy}+v^3g_{yyy})/3 - \ldots + \qquad (A7.b)$$

And third degrees terms of Eq. (A7) have to be identified to those ones of Eq. (A4).

## References

[1] J. Eggers, "Sand as a Maxwell demon", *Phys. Rev. Lett.* **83**, 5322-25, (1999); J. Javier Brey, F. Moreno, R. Garcıa-Rojo and M. J. Ruiz-Montero, "Hydrodynamic Maxwell Demon in granular systems", *Phys. Rev.* **E 65**, p. 11305 (2001)

[2] P. Jean, H. Bellenger, P. Burban, L. Ponson & P. Evesque, "Phase transition or Maxwell's demon in Granular gas?", *Poudres & Grains* **13** (3), 27-38 (juillet-Août 2002), http://www.mssmat.ecp.fr/html_petg/rubrique.php3?id_rubrique=1

[3] R. Mikkelsen, K. van der Weele, D. van der Meer, M. van Hecke and D. Lohse, "Small-number statistics near the clustering transition in a compartmentalized granular gas", *Phys. Rev.* **E 71**, p. 41302 (2005).

[4] M. Leconte, P. Evesque, "Maxwell demon in Granular gas: a new kind of bifurcation? The hypercritical bifurcation", arXive: physics/0609204 (Sept 2006), http://arxiv.org/PS_cache/physics/pdf/0609/0609204.pdf ;

[4bis] P. Evesque, " How one can make the bifurcation of Maxwell's demon in Granular Gas Hyper-Critical ",*Poudres & Grains* **16** (1), 1-20 (Février 2007), http://www.mssmat.ecp.fr/html_petg/rubrique.php3?id_rubrique=1

[5] A. Barrat & E. Trizac, "A molecular dynamics "Maxwell Demon" experiment for granular mixtures", ArXive:Cond-mat/0212054v1 (dec 2002)

[6] P. Manneville, Dynamique non linéaire et chaos, cours du CEA de Physique des Liquides, Paris (2002); wilkipedia, http://en.wikipedia.org/wiki/Dynamical_system

[7] We do not discuss the reason of this linearity with N; it is linked probably to some existing "defect" in the experiment which generates horizontal motion of the ball coupled to vertical motion.

[8] P. Evesque, "Are Temperature and other Thermodynamics Variables efficient Concepts for describing Granular Gases and/or Flows? " , *poudres & grains* **13** (2), 20-26 (2 003) http://www.mssmat.ecp.fr/html_petg/rubrique.php3?id_rubrique=1

[9] M. Hou, "Oscillation in "Temperature" in a compartmentalized Bidisperse granular gas", preprint (2006)

[10] D. van der Meer, K. van der Weele, and D. Lohse, "Bifurcation diagram for compartmentalized granular gases", *Phys. Rev.* **E 63**, 061304 (2001)

[11] D. van der Meer, K. van der Weele & P. Reimann, "Granular fountains: Convection cascade in a compartmentalized granular gas", *Phys. Rev.* **E 73**, 061304 (2006)